\documentclass[twocolumn,aps,prl,showpacs,amsmath,superscriptaddress,longbibliography,notitlepage]{revtex4-1}
\usepackage{amssymb}
\usepackage{mathrsfs}
\usepackage{graphicx}
\usepackage{float}
\usepackage[caption=false]{subfig}
\usepackage[pdftex,colorlinks=red]{hyperref}
\usepackage{verbatim}
\usepackage{txfonts}
\usepackage{epstopdf}
\usepackage[normalem]{ulem}
\usepackage{xcolor}
\usepackage{bm}
\usepackage[utf8]{inputenc}

\begin{document}
\title{Direction reversal of non-Hermitian skin effect via coherent coupling}
\author{Linhu Li}\email{lilh56@mail.sysu.edu.cn}
\affiliation{Guangdong Provincial Key Laboratory of Quantum Metrology and Sensing $\&$ School of Physics and Astronomy, Sun Yat-Sen University (Zhuhai Campus), Zhuhai 519082, China}
\author{Wei Xin Teo}
\affiliation{Department of Physics, National University of Singapore, Singapore 117551, Republic of Singapore}
\author{Sen Mu}\email{senmu@u.nus.edu}
\affiliation{Department of Physics, National University of Singapore, Singapore 117551, Republic of Singapore}
\author{Jiangbin Gong}\email{phygj@nus.edu.sg}
\affiliation{Department of Physics, National University of Singapore, Singapore 117551, Republic of Singapore}
\date{\today}

\begin{abstract}
{Absolute negative mobility (ANM) in nonequilibrium systems depicts the possibility of particles propagating toward the opposite direction of an external force.  We uncover in this work a phenomenon analogous to ANM regarding eigenstate localization and particle transport in non-Hermitian systems under the influence of the non-Hermitian skin effect (NHSE).   
A coherent coupling between two non-Hermitian chains individually possessing the same preferred direction of NHSE is shown to cause a direction reversal of NHSE for all eigenmodes.  This concept is further investigated in terms of time evolution dynamics using a non-Hermitian quantum walk platform within reach of current experiments.  Our findings are explained both qualitatively and quantitatively. The  possible direction reversal of NHSE can potentially lead to interesting applications.}
\end{abstract}

\maketitle

{\it Introduction.-}
%In quantum mechanics, Hermiticity of Hamiltonians is usually assumed to ensure the conservation of probability and real eigen-energies.
%On the other hand,
{Non-Hermitian Hamiltonians provide an effective description of open quantum systems or wave systems with gain and loss \cite{Bender1998nonH,bender2007making,Rotter2009non,ashida2020non}.
%Recently, there has been a growing interest in such systems as they exhibit various intriguing features unique in non-Hermitian systems only, e.g. the exceptional degeneracies \blue{[cite]}, the winding topology of complex spectra \cite{[cite]},
One main feature of non-Hermitian lattice systems with nonreciprocity is the seminal non-Hermitian skin effect (NHSE) under open boundary conditions~\cite{alvarez2018non,yao2018edge,yokomizo2019non,Lee2019anatomy}.  NHSE causes directional accumulation of eigenmodes at the system's boundaries and has a rather deep connection with the point-gap topology of the complex spectrum of non-Hermitian systems
\cite{alvarez2018non,yao2018edge,Yao2018nonH2D, yokomizo2019non,Lee2019anatomy,song2019non, borgnia2020nonH,okuma2020topological,zhang2019correspondence,Lee2019hybrid,li2020topological,li2020critical,yi2020nonH}.
NHSE has spurred considerable interest in condensed matter physics research because it challenged our conventional thinking of bulk-edge correspondence and has motivated the so-called non-Bloch band theory \cite{yao2018edge,yokomizo2019non,Lee2019anatomy}.  Much attention has also been paid to the
interplay between the NHSE and other important physical effects, such as topological localization \cite{Lee2019hybrid,li2020topological}, external electromagnetic fields \cite{ming2021magnetic,deng2021non,peng2022manipulating}, disorders and defects \cite{jiang2019interplay,longhi2019topological,zeng2020topological,li2021impurity,guo2021exact,liu2021exact2,bhargave2021non,schindler2021dislocation,sun2021geometric}}. %and even between different NHSE channels in the same system \cite{li2022non}.
%how the NHSE manifests in various scenarios
%e.g.
%hybridization with topological localization \cite{Lee2019hybrid,li2020topological},
%manipulation by external electromagnetic fields \cite{ming2021magnetic,deng2021non,peng2022manipulating},
%interplay with disorders and defects in 1D \cite{jiang2019interplay,longhi2019topological,zeng2020topological,li2021impurity,guo2021exact,liu2021exact2}
%and two dimenional (2D) systems \cite{bhargave2021non,schindler2021dislocation,sun2021geometric},
%and non-Hermitian pseudo-gaps emerging from competition between different NHSE channels \cite{li2022non}.

Non-reciprocal hopping on a one-dimensional (1D) lattice defines a preferred direction analogous to a physical direction of an external force.  The preferred boundary for bulk state localization as NHSE is thus intuitive, so does the preferred direction favoring particle transport \cite{song2019non,yi2020nonH,li2020topological,Wanjura2020,wanjura2021correspondence,xiao2020non,xiao2021observation,wang2021detecting}.
For example, if the strength of intercell hopping to the left is always larger than that to the right, then NHSE is expected to localize all states at the left boundary. On the other hand, we must take note of one remarkable dynamical phenomenon, namely, absolute negative mobility (ANM), where particles propagate toward the opposite direction of an external force. Seemingly contradictory to Newton’s second law, ANM has already been
widely investigated and experimentally realized in various systems far from equilibrium
\cite{keay1995dynamic_ANM,cannon2000absolute_ANM,
eichhorn2002brownian_ANM,eichhorn2002paradoxical_ANM,
machura2007absolute_ANM,nagel2008observation_ANM,%Josephson junction
ros2005absolute_ANM,
reguera2012entropic_ANM,slapik2019tunable_ANM,%particle separation
ghosh2014giant_ANM,sarracino2016nonlinear_ANM}.  Recognizing non-Hermitian systems as nonequilibrium systems,  it is necessary to address the possibility of population accumulation or particle transport in a direction against the preferred direction indicated by the non-reciprocal hopping.
This issue is not only of theoretical interest, but may offer versatile control knobs to manipulate NHSE for various applications, such as light funneling \cite{weidemann2020topological} and directional signal amplification \cite{Wanjura2020,wanjura2021correspondence,xue2021simple}.

{In this work, we unveil a general scheme to induce 1D NHSE in a direction precisely opposite to the favored direction of non-reciprocal hopping.  As shown below, this exotic phenomenon can be obtained at both the eigenstate level and the dynamics level.  There can be multiple interpretations of why a direction reversal of NHSE occurs. Among them,  a simple physical picture adopted below is based on the interference between multiple hopping pathways.  Specifically, if two non-Hermitian lattices with the same preferred non-reciprocal direction are coupled,  then multiple hopping pathways become available.  The resulting interference between the multiple hopping pathways can counter-intuitively and drastically alter the effective strengths of hopping towards two directions, and hence one must reexamine the true physically favored direction of NHSE. }

{The direction reversal of NHSE by coherent coupling is in principle observable in a variety of quantum and classical platforms realizing NHSE \cite{brandenbourger2019non,helbig2020generalized,ghatak2020observation,weidemann2020topological,hofmann2020reciprocal,xiao2020non,xiao2021observation,wang2021detecting,liang2022observation}.
In particular, reversed NHSE at the eigenstate level is already within the reach of classical platforms, such as circuits. However, how reversed NHSE is manifested at the dynamics level is less straightforward. We hence propose a non-unitary quantum walk setting directly addressing non-Hermitian dynamics \cite{xiao2020non,xiao2021observation,wang2021detecting}, with the essential addition being an interchain hopping for the quantum walker along two chains. As elaborated below, even though the preferred direction of NHSE is no longer obvious in the quantum walk dynamics, the multiple propagation pathways induced by the interchain hopping can still lead to a direction reversal of particle transport. }
%a non-Hermitian absolute negative mobility (NHANM) as a manifestation of the reversed non-reciprocal accumulation.
%In a non-unitary two-chain quantum walk, hybridization between the two chains can be realized through a local interchain unitary shift, and the resultant reversed non-reciprocal accumulation manifests as a non-Hermitian absolute negative mobility (NHANM) in the dynamical evolution.
%Such behavior may also prove useful in manipulating the NHSE in its various applications, such as light funneling \cite{weidemann2020topological} and directional signal %amplification \cite{Wanjura2020,wanjura2021correspondence,xue2021simple}.

\begin{figure}
\includegraphics[width=0.8\linewidth]{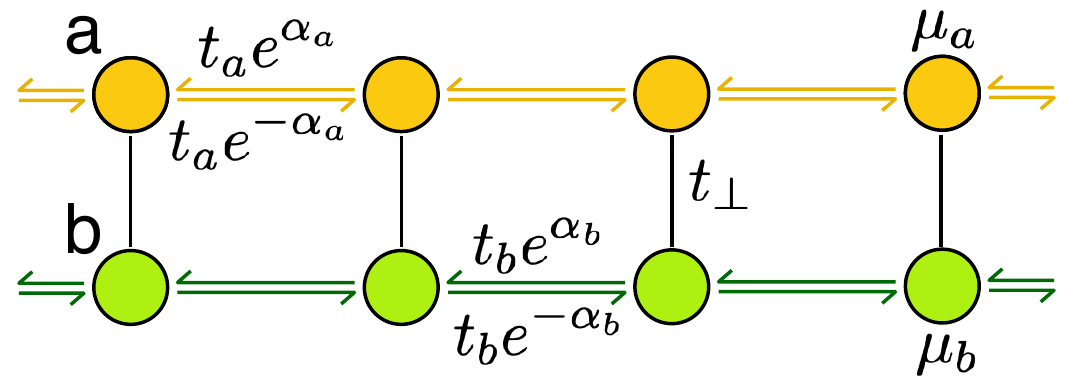}
\caption{Schematic of two coupled chains with non-reciprocal hoppings.  In either lattice, the amplitude of hopping to the left direction is $t_s e^{\alpha_s}$ ($s=a, b)$, apparently with a magnitude larger than that to the right direction $t_s e^{-\alpha_s}$.  $t_\perp$ introduces coupling between the two chains and hence multiple hopping pathways from one lattice site to its neighboring site. }
\label{fig:model}
\end{figure}

\begin{figure}
\includegraphics[width=1\linewidth]{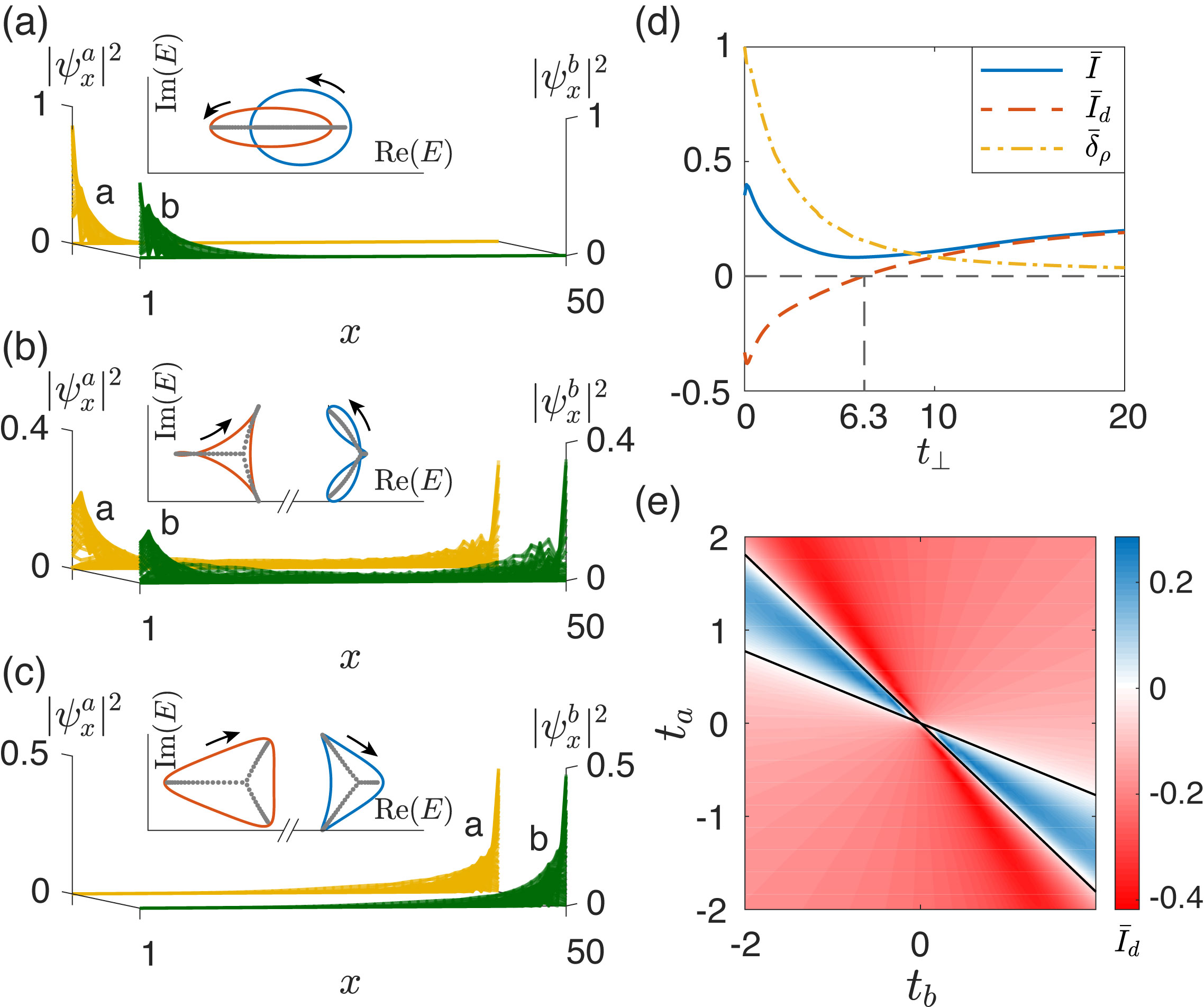}
\caption{
(a-c) Distribution of all eigenmodes on the two chains, with different interchain coupling $t_\perp=0,6,15$ respectively. Insets show the corresponding spectra under PBC (red and blue for the two bands) and OBC (gray). 
{Clockwise and anti-clockwise winding directions of the PBC spectra versus the quasi-momentum $k$, as indicated by the black arrows, correspond to OBC skin modes localized on the left and right, respectively \cite{borgnia2020nonH,zhang2019correspondence,okuma2020topological}.}
%The OBC spectra fall within the areas enclosed by the PBC loop, indicating a skin localization of the eigenmodes under the OBC \cite{borgnia2020nonH,zhang2019correspondence,okuma2020topological}.
Note that in (b) and (c), we have omitted a large spacing in ${\rm Re}[E]$ between the two energy bands (red and blue), represented by the double slash in the axis. Other parameters are $t_a=0.75$, $t_b=-1$, $\alpha_a=0.5$, $\alpha_b=0.2$, and $\mu_a=-\mu_b=0.5$.
(d) $\bar{I}$, $\bar{I}_d$, and $\bar{\delta}_\rho$ defined in Eqs. \eqref{IPR}, \eqref{dIPR} and \eqref{distribution_diff},
versus the interchain coupling $t_\perp$, with the same parameters as in (a-c).
(e) Phase diagram regarding the directional IPR $\bar{I}_d$ at $t_{\perp}=30$ with other parameters the same as (a-c).
Black lines are obtained from the perturbation results of Eq. \eqref{eq:boundary}.
%Positive and negative values $\kappa$ correspond to opposite directions of the non-reciprocal accumulation, with the phase boundaries at $\kappa=0$ (black lines) given by Eq. \eqref{eq:boundary}.
}
\label{fig:fig2}
\end{figure}

{\it Direction reversal of eigenstate population accumulation.-}
{Our starting point is a minimal model depicting two coupled non-Hermitian chains \cite{HN1996prl} with different on-site potentials, as shown in Fig.~\ref{fig:model}.
The real-space Hamiltonian reads
\begin{eqnarray}
\hat{H} &=&\sum_{x=1}^L \sum_{s=a,b}\left[t_se^{\alpha_s}\hat{s}^\dagger_x\hat{s}_{x+1}+t_s e^{-\alpha_s}\hat{s}^\dagger_x\hat{s}_{x-1}\right.\nonumber\\
%\hat{H} &=&\sum_x^L \left[t_ae^{\alpha_a}\hat{a}^\dagger_x\hat{a}_{x+1}+t_a e^{-\alpha_a}\hat{a}^\dagger_x\hat{a}_{x-1}+t_be^{\alpha_b}\hat{b}^\dagger_x\hat{b}_{x+1}+t_be^{-\alpha_b}\hat{b}^\dagger_x\hat{b}_{x-1}\right.\nonumber\\
&&\left.+t_{\perp}(\hat{a}^\dagger_x\hat{b}_{x}+\hat{b}^\dagger_x\hat{a}_{x})+\mu_a\hat{a}^\dagger_x\hat{a}_x+\mu_b\hat{b}^\dagger_x\hat{b}_x,\right]\label{eq:ladder}
\end{eqnarray}
with $t_{s}$ and $\alpha_{s}>0$ determining the asymmetric hopping amplitudes on the two chains labeled by $s=a,b$.  Referring to Fig.~\ref{fig:model}, the preferred direction of the non-reciprocal hopping here is apparently to the left for both chains.  
The on-site potential is set to be $\mu_a=-\mu_b=\mu$, with all other choices being equivalent upon shifting their eigenenergies. The two chains are completely decoupled if $t_\perp=0$, each displaying NHSE localization at the left edge,  with an inverse localization length $\kappa_{a,b}=\alpha_{a,b}$\cite{yao2018edge,Lee2019anatomy,Lee2019hybrid}. 
%For a reason to be explained later,  the NHSE localization lengths $\alpha_{a}$ and $\alpha_{b}$ are assumed to be different.
An example depicting such a decoupling limit is illustrated in Fig.~\ref{fig:fig2}(a).}
% where the nonreciprocal population accumulation in chain $a$ chain is seen to be stronger than that in chain $b$.}

{Upon switching on the interchain coupling ($t_{\perp}\ne 0)$, both the complex spectrum and eigenstate localization of the coupled system start to differ from that of the uncoupled case, no matter how small $t_{\perp}$ is 
 \cite{li2020critical,liang2021anomalous,mu2021nonhermitian}.   To allow for many hopping pathways from one site to its neighboring site,  such as  $a_x\rightarrow  a_{x+1}$ and $a_x\rightarrow b_x\rightarrow b_{x+1}\rightarrow a_{x+1}$, to interfere significantly, here 
we investigate a strong coupling regime with sufficiently large  $t_{\perp}$.  
It is then found that {\it all} eigenmodes can now localize at the opposite edge as compared with that in the uncoupled case. This is clearly seen in 
Fig.~\ref{fig:fig2}(a)-(c) as the interchain coupling strength $t_{\perp}$ increases from $0$ to $6$ and to $15$.}
%This model recovers the two-chain model in the Critical NHSE paper when $t_a=t_b$ and $\alpha_a=-\alpha_b$, or the one in the mobility edge paper when $t_a=t_b$ and $\alpha_b=0$.

{To characterize the above-observed direction reversal of NHSE, we consider averages of the standard and directional inverse participation ratios (IPR and dIPR), defined as
\begin{eqnarray}
\bar{I}&=&\frac{1}{2L}\sum_m\sum_{x=1}^L\left(|\psi^a_{x,m}|^4+|\psi^b_{x,m}|^4\right),\label{IPR}\\
\bar{I}_{d}&=&\frac{1}{2L}\sum_m\sum_{x=1}^L\frac{\left(x-(L+1)/2\right)\left(|\psi^a_{x,m}|^4+|\psi^b_{x,m}|^4\right)}{(L-1)/2},\label{dIPR}
\end{eqnarray}
with $\psi_{x,m}^s$ the wave amplitude of the $m$-th normalized right eigenmode at site $x$ of sublattice $s$.
Representative results are presented in Fig.~\ref{fig:fig2}(d).
It is seen that the IPR (and the absolute value of dIPR) gets larger either for weaker or stronger $t_\perp$, indicating a stronger boundary accumulation of the eigenmodes, but with opposite accumulating directions, as evidenced by the signs of the dIPR.
A reversal of the NHSE direction starts to occur when $\bar{I}_d=0$, which is at $t_\perp\approx 6.3$ in the shown example.   In the neighborhood of the transition point $\bar{I}_d=0$ [see Fig.~\ref{fig:fig2}], the eigenmodes can possibly localize at both boundaries in a balanced manner as the bipolar NHSE \cite{song2019realspace,zhang2021acoustic}.  More importantly, away from the transition point,  {\it all} eigenmodes are now localized at the opposite boundary as compared with the uncoupled case. 
Meanwhile, the difference between the average distribution on the two lattices, defined as
\begin{eqnarray}
\bar{\delta}_\rho=\frac{1}{2L}\sum_m\big|\sum_{x=1}^L\left(|\psi^a_{x,m}|^2-|\psi^b_{x,m}|^2\right)\big|,\label{distribution_diff}
\end{eqnarray}
is seen to decrease with increasing $t_\perp$, indicating a stronger hybridization between the two chains at larger $t_\perp$.}

%that the reversal of NHSE direction does not necessarily coincide with the occurence of a minimal of $\bar{I}$ (i.e. where the eigenmodes are maximumly delocalized), 
%Indeed, $\bar{I}_d$ only indicates the average accumulaton direction over all eigenmodes,
%yet near the transition point of NHSE direction, different eigenmodes are seen to accumulate to different lattice ends [see Fig.~\ref{fig:fig2}(b)], the so-called bipolar NHSE observed before \cite{song2019realspace,zhang2021acoustic}.}
%That is, in the coupled NH chains, this transition determined by $\bar{I}_d=0$ reflects a balanced non-reciprocal accumulation toward both %directions for the overall system.
%Nevertheless, the emergence of the bipolar NHSE also depends on specific models, and does not always accompany the transition between normal and reversed non-reciprocal accumulations. An example is the quantum walk system discussed latter in this paper, where the complex spectral winding vanishes at the transition point, corresponding to disappearance of the NHSE.

{{\it Physics of reversed NHSE.-}
To confirm that the main physics behind reversed NHSE is the interference between multiple hopping pathways,
we consider a straightforward first-order perturbation theory by treating $\hat{H}_\perp=t_{\perp}\sum_{x}(\hat{a}^\dagger_x\hat{b}_{x}+\hat{b}^\dagger_x\hat{a}_{x})$ as the unperturbed Hamiltonian. The unperturbed eigenmodes at site $x$ are simply given by local hybridized adiabatic modes of the coupled system, i.e.,
\begin{eqnarray}
|u_{\pm,x}\rangle=\hat{u}^\dagger_{\pm,x}|0\rangle=(\hat{a}_x^\dagger\pm\hat{b}_x^\dagger)|0\rangle/\sqrt{2},~E_{\pm,x}=\pm t_\perp,
\end{eqnarray}
with $|0\rangle$ the vacuum and $E_{\pm,x}$ the corresponding unperturbed eigenenergies due to the coherent coupling.  By taking all the rest terms as a perturbation, we rewrite the non-reciprocal hopping Hamiltonian in the local adiabatic representation, yielding
\begin{eqnarray}
\hat{H}'_{\pm}&=&\sum_{x}(t_ae^{\alpha_a}+t_be^{\alpha_b})\hat{u}^\dagger_{\pm,x}\hat{u}_{\pm,x+1}+
(t_ae^{-\alpha_a}+t_be^{-\alpha_b})\hat{u}^\dagger_{\pm,x+1}\hat{u}_{\pm,x}\nonumber\\
&&+(\mu_a\pm\mu_b)\hat{u}^\dagger_{\pm,x}\hat{u}_{\pm,x}.\label{eq:H_perturbed}
\end{eqnarray}
Interestingly, other than the local on-site energy being expectedly different, the ``+" and ``-" hybridized lattice sites have the same non-reciprocal hopping strengths, $(t_ae^{\alpha_a}+t_be^{\alpha_b})$ to the left and $(t_ae^{-\alpha_a}+t_be^{-\alpha_b})$ to the right.  That is,  the effective hopping amplitudes are seen to be a sum of two individual hopping amplitudes $t_ae^{\alpha_a}$ and $t_be^{\alpha_b}$ 
or $t_ae^{-\alpha_a}$ and $t_be^{-\alpha_b}$,  thus clearly indicating an interference mechanism.  Most importantly, if $t_a$ and $t_b$ are of different signs, then there is a destructive interference between the two favored amplitudes.  This can then lead to 
\begin{eqnarray}
\big|t_a e^{-\alpha_a}+t_be^{-\alpha_b}\big|
>
\big|t_a e^{\alpha_a}+t_be^{\alpha_b}\big|, \label{eq:phase_boundary}
\end{eqnarray}
meaning that NHSE here should accumulate/localizate population to the right for {\it all} the eigenmodes,  opposite to the NHSE direction on the uncoupled chains. 
Inequality \eqref{eq:phase_boundary} also suggests that reversed NHSE occurs within the following parameter regime
\begin{eqnarray}
\frac{t_a}{-t_b}\in(\frac{e^{\alpha_b}-e^{-\alpha_b}}{e^{\alpha_a}-e^{-\alpha_a}},\frac{e^{\alpha_b}+e^{-\alpha_b}}{e^{\alpha_a}+e^{-\alpha_a}}),\label{eq:boundary}
\end{eqnarray}
as shown by the solid lines in Fig.~\ref{fig:fig2}(e).  The transition lines obtained this way match well with the numerical results based on the sign of the average directional IPR $\bar{I}_d$.   A  momentum space perturbation theory
together with the so-called generalized Brillouin zone \cite{yao2018edge,yokomizo2019non} yields the same prediction in theory, as detailed in Supplementary Materials. Note also that the role of the on-site potential difference $\mu$ is not seen here due to our first-order treatment or the strong coupling assumption.  The actual threshold value $t_\perp$ to enter the reversed NHSE regime gradually increases with $\mu$.   Reversed NHSE may also be obtained under $t_a t_b>0$, if we introduce multiple hopping pathways in other manners, such as allowing for off-diagonal couplings between the two chains.  These details can be found in Supplementary Materials.}

{\it Reversed particle transport in non-Hermitian quantum walk.-}
\begin{figure}
\includegraphics[width=1\linewidth]{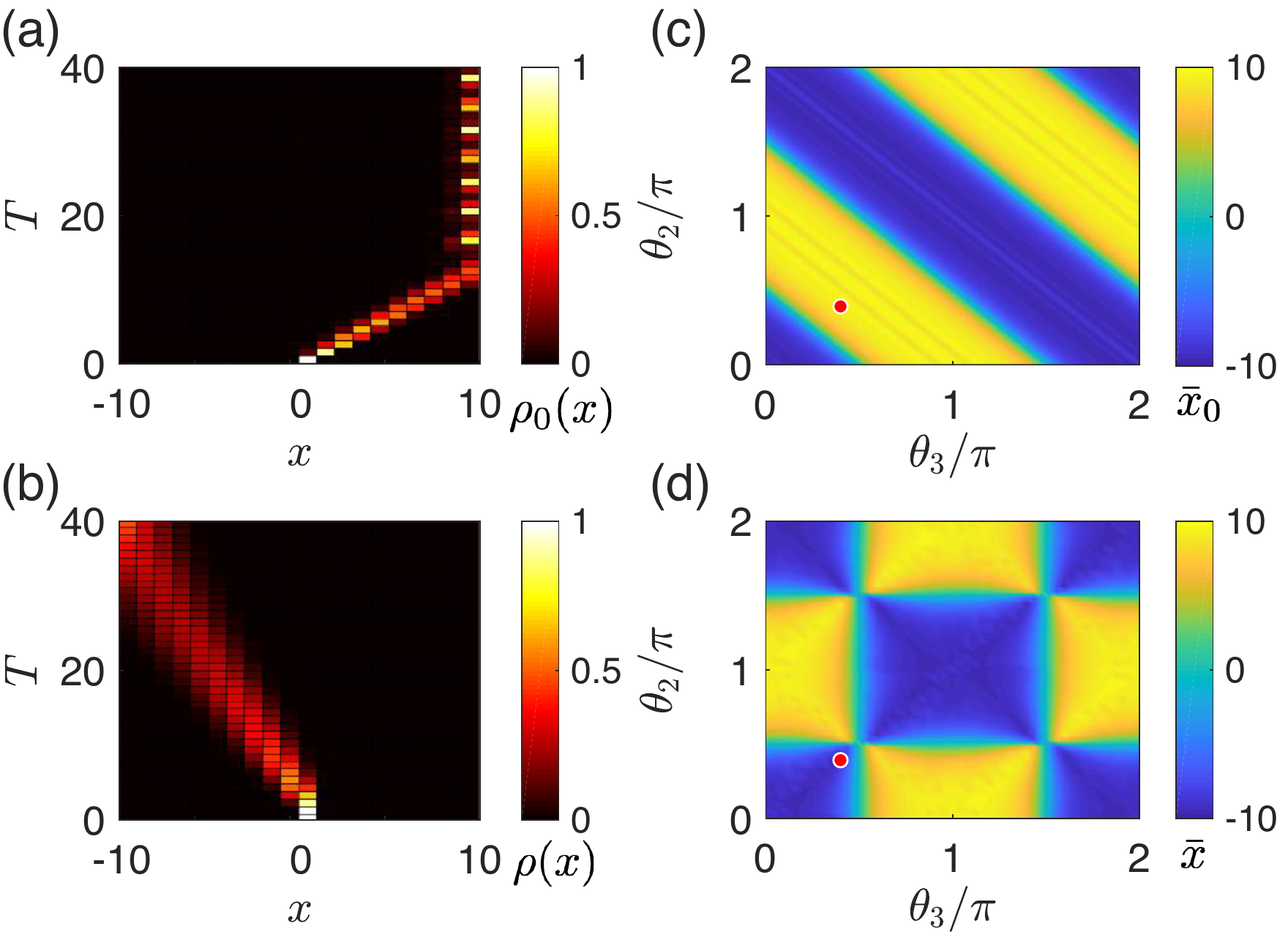}
\caption{
(a,b) Spatial distributions of the quantum walks for an initial state prepared in the middle of the system, governed by $U_0$ and $U$, the two quantum walks without and with the interchain hopping, respectively.
(c,d) Average position of the final for the two quantum walks at $T=40$, versus the two angles $\theta_2$ and $\theta_3$. Yellow and blue (bright and dark) regimes indicate non-reciprocal pumping toward the directions of $x=N$ and $x=-N$ respectively. The red points in (c,d) represents the two cases of (a,c), with $\theta_2=\theta_3=0.4\pi$. Other parameters are $\alpha_a=\alpha_b=3$, $\theta_1=0.2\pi$ and $N=10$.
}
\label{fig:QW}
\end{figure}
{So far, the reversed NHSE is investigated on the eigenstate level via population accumulation against the preferred direction of non-reciprocal hopping.  To make a closer analogy to ANM, it is necessary to explore how this leads to particle transport along a reversed direction. 
%This NHANM can be easily seen by considering a time-evolution of the above system, and the propagating direction can be intuitively extracted from the asymmetric couplings in the effective Hamiltonian of Eq. \eqref{eq:H_perturbed}.
To motivate experimental interest, we use an available and fruitful platform, namely, 
 a discrete-time non-unitary quantum walk model realizing the NHSE through single-photon dynamics in a 1D chain \cite{xiao2020non,xiao2021observation,wang2021detecting}.  We now propose a quantum walker along two chains, plus a local interchain exchange depending on the spin state.}

{Specifically, we consider the following two Floquet operators governing the quantum walk, 
%\red{[I think we may need to keep these details since in Figs. (3,4) we use parameters $\theta_{2,3}$ to show a phase diagram of the %NHANM.]}
%Here we further consider an alternative example which lacks a simple Hamiltonian description,
%namely discrete-time non-unitary quantum walk on two coupled 1D chains, governed by the Floquet operator
\begin{eqnarray}
U_0&=&R(\theta_1) S_2 R(\theta_2+\theta_3) M  R(\theta_2+\theta_3) S_1 R(\theta_1),\label{eq:floquet0}\\
U&=&R(\theta_1) S_2 R(\theta_2) S_4 R(\theta_3) M  R(\theta_3) S_3R(\theta_2) S_1 R(\theta_1).
\label{eq:floquet}
\end{eqnarray}
Here $R(\theta)$ rotates the spin by $\theta$ about the $y$ axis, with
$R(\theta)=\sum_{x=-N}^N\sum_{s=a,b}|s,x\rangle\langle s,x|\otimes e^{-i\lambda_s\theta\sigma_y/2}$,
$s=a,b$ denoting the two chains, $x$ the site index, and $\lambda_a=1$ and $\lambda_b=-1$.
The shift operators $S_1$ and $S_2$ are standard quantum walk operations, as they shift the walker to the left and right along either chain, if and only if the spin is up and down, respectively. 
% $|\uparrow\rangle$ and $|\downarrow\rangle$ along $x$ direction respectively,
%\begin{eqnarray}
%S_1&=&\sum_{x=-N}^N\sum_{s=a,b}\left(|s,x\rangle\langle s,x|\otimes|\downarrow\rangle\langle \downarrow|+|s,x+1\rangle\langle s,x|%\otimes|\uparrow\rangle\langle \uparrow|\right),\nonumber\\
%\end{eqnarray}
%\begin{eqnarray}
%S_2&=&\sum_{x=-N}^N\sum_{s=a,b}\left(|s,x-1\rangle\langle s,x|\otimes|\downarrow\rangle\langle \downarrow|+|s,x\rangle\langle s,x|%\otimes|\uparrow\rangle\langle \uparrow|\right).\nonumber\\
%\end{eqnarray}
Non-unitarity/non-Hermiticity is introduced through the operator $M$, with
$$M=\sum_{x=-N}^N\sum_{s=a,b}|s,x\rangle\langle s,x|\otimes\left(|\downarrow\rangle\langle\downarrow|+e^{-\alpha_s}|\uparrow\rangle\langle\uparrow|\right),$$
describing the (quasi-)particle loss only for the spin-up component. $U_0$ thus defined above yields exactly two copies of the quantum walk model realizing the NHSE in Refs.~\cite{xiao2020non,xiao2021observation,wang2021detecting}, but with opposite spin rotation angles  through the parameter $\lambda_s$.  The $M$ operator alone seems to suggest that the spin-down channel is favored. This effect further interplays with the spin rotation operator $R(\theta)$ and the shift operators $S_{1,2}$ to yield non-reciprocal particle transport, with the preferred direction no longer obvious. Despite the difference in $\lambda_s$ between the two quantum walk copies, their preferred direction of transport is found to be always the same. }
 
\begin{figure}
\includegraphics[width=1\linewidth]{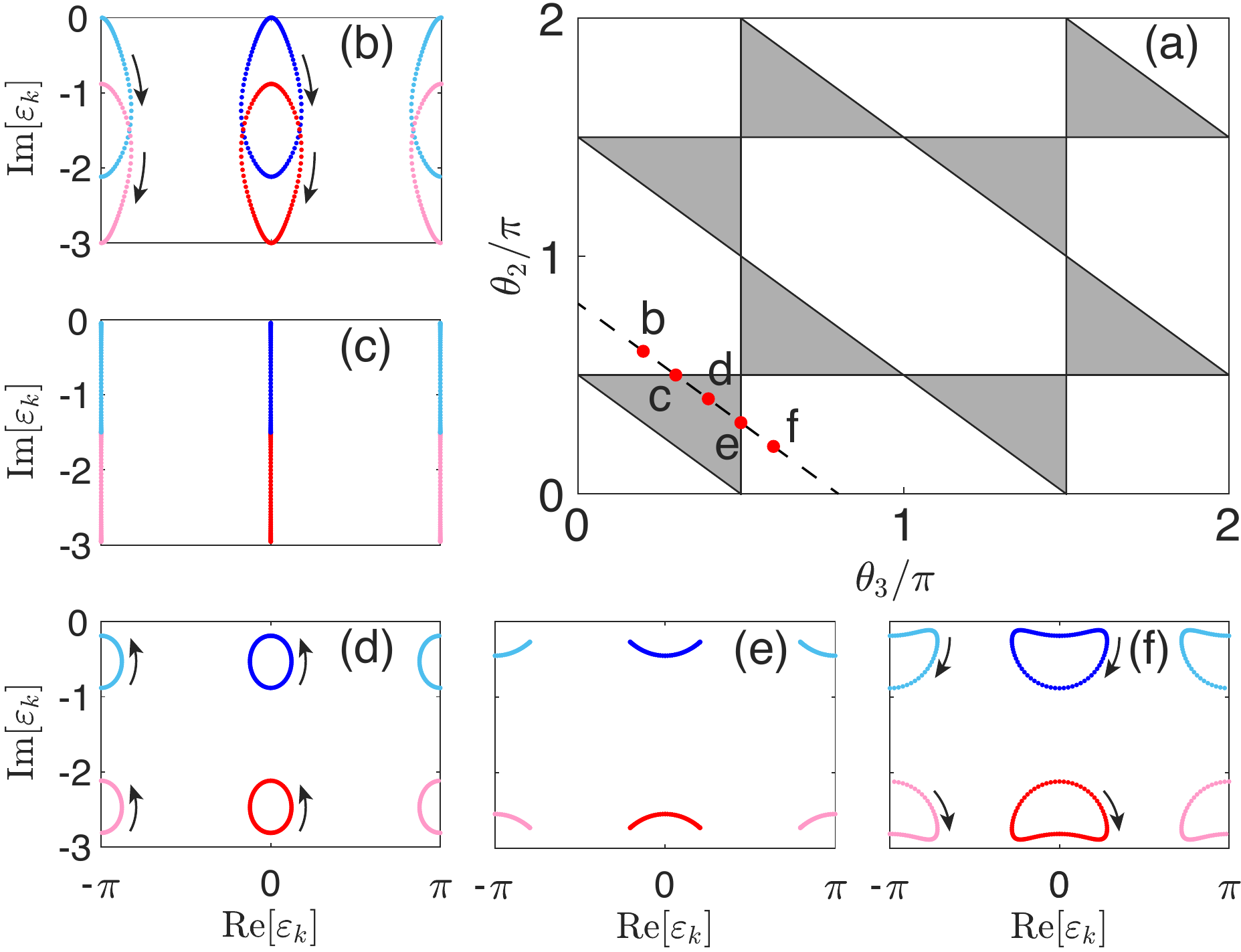}
\caption{
(a) A phase diagram obtained from Fig. \ref{fig:QW}(c) and (d), reversed non-reciprocal accumulation occurs for the parameters falling in the gray areas.
(b) to (f) The quasi-energy spectra of $U_k$ with different parameters, corresponding to the five red dots along the black dash line with $\theta_2+\theta_3=0.8\pi$ in panel (a).
Different colors mark the four bands of the spectra (blue, red, cyan, and pink).
Black arrows indicate the winding direction of the quasi-energies with $k$ varying from $0$ to $2\pi$.
The parameters are $\theta_2=0.2\pi$, $0.3\pi$, $0.4\pi$, $0.5\pi$, and $0.6\pi$ from (b) to (f),
with $\alpha_a=\alpha_b=3$ and $\theta_1=0.2$ for all panels.
}
\label{fig:QW_phase}
\end{figure}

{We now couple the two chains accommodating $U_0$, thus defining our quantum walk model $U$.  $U$ is obtained by inserting $S_3$ and $S_4$ into $U_0$. $S_3$ and $S_4$ are almost the same operations as
$S_1$ and $S_2$ except that the walker is instructed to hop onto the other chain (of the same lattice index) when the spin state is up and down, respectively. Detailed definitions of these operations are shown in Supplementary Materials.   $S_3$ and $S_4$ are expected to hybridize the two chains and induce interference between multiple hopping pathways.  We aim to show that even though the two individual chains have the same preferred walk direction, their coupling can reverse the direction of transport, thus demonstrating direction reversal of NHSE via time evolution dynamics. }
{We consider an initial state prepared in the middle of the system, $\Psi_{\rm ini}=\frac{1}{\sqrt{2}}(|a,0\rangle\otimes|\uparrow\rangle+|b,0\rangle\otimes|\uparrow\rangle).$ In Fig.~\ref{fig:QW}(a) and (b), we show the spatial distribution $\rho_0(x)$ of the normalized final state $\Psi_{0,{\rm fin}}=U_0^T\Psi_{\rm ini}$  and $\rho(x)$ of $\Psi_{\rm fin}=U^T\Psi_{\rm ini}$ for the quantum walk governed by $U_0$ and $U$, respectively.  
Here $T$ represents the number of steps of the quantum walk, and the normalized spatial distribution is defined as
$\rho(x)=\sum_{s=a,b,\sigma=\uparrow,\downarrow}|\tilde{\psi}^{x,s,\sigma}_{\rm fin}|^2,$ with $\tilde{\psi}^{x,s,\sigma}_{\rm fin}$ the wave amplitude of the normalized final state,
\begin{eqnarray}
\tilde{\psi}^{x,s,\sigma}_{\rm fin}&=&\frac{\psi^{x,s,\sigma}_{\rm fin}|s,x\rangle\otimes|\sigma\rangle}{\sqrt{\sum_{x,s=a,b,\sigma=\uparrow,\downarrow}|\psi^{x,s,\sigma}_{\rm fin}|^2}},\nonumber
\end{eqnarray}
obtained from $\Psi_{\rm fin}={\sum_{x,s=a,b,\sigma=\uparrow,\downarrow}\psi^{x,s,\sigma}_{\rm fin}|s,x\rangle\otimes|\sigma\rangle}.$
It is clearly seen from Fig.~\ref{fig:QW}(a) and (b) that introducing the interchain hopping reverses the propagation direction of the walker.
%Note that the interchain shift in a quantum walk is represented by a unitary matrix, therefore it doesn't make sense to say it is strong or %weak, as for the interchain couplings in the static model.
%However, numerically we find that eigenmodes of the Floquet operator $U$ are equally distributed on the two chains,
%analogous to the coupled HN chains in the strongly hybridized regime.
In Fig.~\ref{fig:QW}(c) and (d), we further examine the average position of the final state, defined as
%$$\bar{x}=\sum_{x,s=a,b,\sigma=\uparrow,\downarrow}|\psi^{x,s,\sigma}_{\rm fin}|^2[x-(N+1)/2].$$
$\bar{x}=\sum_{x,s=a,b,\sigma=\uparrow,\downarrow}x\ |\psi^{x,s,\sigma}_{\rm fin}|^2.$  Note that this average is over both chains.
Without the interchain hopping operators $S_{3,4}$, the quantum walk governed by $U_0$ exhibits the NHSE, of which the direction of non-reciprocal population accumulation is determined by the two rotation angles $\theta_1$ and $\theta_2+\theta_3$ in Eq.~\eqref{eq:floquet0}. As seen in Fig.~\ref{fig:QW}(c) and (d), the interchain hopping can reverse the direction of particle transport.  That is, when the color in  Fig.~\ref{fig:QW}(c)  mismatches that in Fig.~\ref{fig:QW}(d), reversed particle transport, as compared with the decoupled case, occurs.   
Combining the results in Fig. \ref{fig:QW}(c) and (d), we obtain the parameter regime in Fig.~\ref{fig:QW_phase}(a) on the $\theta_3-\theta_2$ plane, where reversed particle transport is observed.}

%In the previous static model of coupled HN chains, the reversed non-reciprocal accumulation can be directly seen from the spatially asymmetric couplings in the perturbed Hamitlonians of Eq. \eqref{eq:H_perturbed}.
%\blue{For the two-chain quantum walk, however, the effective Floquet Hamiltonian defined as
%$U=e^{-iH_{\rm eff}}$
%$H_{\rm eff}=i \ln U$
%is generally very complicated in real space.
{To further digest the direction reversal of NHSE,  one may also investigate the winding behavior of the quasi-energy $\varepsilon_k$, obtained from $U_k\Psi_k=e^{-i \varepsilon_k}\Psi_k,$
%$$\varepsilon_k=i\ln U_k,$$
with $U_k$ being the Fourier transform of $U$, $\Psi_k$ the eigenvectors of $U_k$, and $k$ the Bloch momentum reflecting the translational invariance of the quantum walk model.  The winding of the quasi-energy spectrum as $k$ increases from $0$ to $2\pi$  is shown in Fig.~\ref{fig:QW_phase}(b) to (f). The direction of the winding is seen to change when the system parameters $(\theta_2, \theta_3)$ move across the phase boundary identified in Fig.~\ref{fig:QW_phase}(a). There is hence a jump of the spectral winding number between $\pm1$ and $0$, as we go from case (b) to case (f).  In particular, as shown in Fig.~\ref{fig:QW_phase}(c) and (e), along the phase boundary, the quasi-spectrum in the $k$-space does not enclose any area, corresponding to a trivial spectral winding and the absence of NHSE.   These results further verify that the above observed reversal of particle transport direction is due to reversed NHSE. }
%Unlike the previously discussed coupled HN chains displaying a bipolar NHSE in the transition regime, in this quantum walk system there is a clear phase boundary distinguishing the normal and reversed non-reciprocal accumulation, as shown by Fig. \ref{fig:QW_phase}(c) and (e) where the quasi-spectrum in the $k$-space encloses no any area, corresponding to a trivial spectral winding and the absence of non-reciprocal accumulation.

%As seen in the figure, the interchain shifts alter the accumulating direction in certain parameter regimes, manifesting the reversed non-reciprocal accumulation discussed in previous sections.
%\red{[Shall we add a total phase diagram of these ``certain parameter regimes"? And perhaps another panel of the eigenvalues showing the winding? In that case my previous note about spectral winding of Floquet operators can be added to the Supp. Mat..]}

%Finally, we note that while our numerics are carried out in a system with a few tens of sites, the non-reciprocal accumulation in non-Hermitian system is a topologically protected phenomenon, and can already been seen in smaller systems, and is thus feasible in various experimental setups.
%Moreover,

{To conclude, we note that particle transport with a reversed direction, as illustrated in Fig.~\ref{fig:QW}(a) and (b),  can be observed within very few quantum walk steps. The required lattice size can also be small since there is no need to distinguish between bulk sites and edge sites.
In Supplemental Materials, we even add an example where reversed NHSE in our quantum walk system can be observed using only two unit cells.}

{\it Summary.-}
{We have shown that a coherent coupling between  two 1D non-Hermitian chains can lead to direction reversal of NHSE for all the eigenmodes.  We have demonstrated this concept using both the spatial profile of stationary solutions, as well as time evolution dynamics on a quantum walk platform within reach of current experiments.  In our first model, the common and individual direction of NHSE is obvious, as observed from the non-reciprocal hopping on the two individual chains, yet a coupling between them yields a population accumulation along the reversed direction.  This intriguing phenomenon is interpreted in terms of the interference between multiple hopping pathways and explained quantitatively via an adiabatic/hybridized representation.  In our second working model aiming at an experimental proposal to observe reversed NHSE on the dynamics level, two individual chains hosting a quantum walker have the same preferred direction of particle transport, yet an interchain hopping can again reverse the direction of particle transport.  In both models, we have witnessed how a physical phenomenon analogous to ANM may emerge in contexts or experimental platforms of non-Hermitian physics. 
%and the NHANM in a dynamical evolution of a non-unitary two-chain quantum walk,
%whose effective Floquet Hamiltonian are usually too cumbersome to analysis.
%where a simple description of an effective Floquet Hamiltonian is usually difficult to obtain.
%where a simple effective Floquet Hamiltonian for describing such systems is usually absent.
%In both cases, the reversed non-reciprocal accumulation occurs when the two chains are strongly hybridized with each other, i.e. each of their %eigenmodes have almost the same distributions on the two chains.
Our findings should also offer useful schemes to manipulate the NHSE by tuning the coherent coupling between individual subsystems. 
It should be stimulating to extend our findings to higher dimensions, where NHSE becomes a rather universal property of non-Hermitian systems, with the boundary localization behavior of bulk eigenmodes strongly dependent on the system's geometry \cite{zhang2021universal}.}

{\it Acknowledgements.--}
L. L. would like to thank Zhihuang Luo and Zhenhua Yu for helpful discussion.
L. L. acknowledges funding support by the National Natural Science Foundation of China (12104519) and the Guangdong Basic and Applied Basic Research Foundation (2020A1515110773).  J. G. acknowledges support from Singapore National Research Foundation  (Grant No. NRF-NRFI2017-04). 

\begin{comment}
By taking the terms of $t_\perp$ and $t_2$ as the unperturbed Hamiltonian,
the unperturbed eigenstates and eigenenergies are given by
\begin{eqnarray}
\psi_{\pm}(k)&=&\frac{1}{\sqrt{2}}\left(
\begin{array}{c}
1\\
\pm e^{i\theta}
\end{array}
\right),
E^0_{\pm}(k)=\pm\sqrt{t_2^2+t_{\perp}^2+2t_2t_\perp\cos k}\nonumber\\
\end{eqnarray}
with $\cos \theta=d_x/\sqrt{d_x^2+d_y^2}$, $d_x=t_\perp+t_2\cos k$ and $d_y=t_2\sin k$.
The first-order perturbation of eigenenergies,
given by
\begin{eqnarray}
\delta E_\pm&=&\langle\psi_\pm |h'| \psi_\pm\rangle\nonumber\\
&=&t_a\cos(k-i\alpha_a)+t_b\cos(k-i\alpha_b),
\end{eqnarray}
is independent from the exact value of $t_\perp$ or $t_2$. That is, it is the same as the previous case with only $t_\perp$. Yet the GBZ solution cannot be obtained in the same way since now the unperturbed eigenenergies $E^0_{\pm}(k)$ are also $k$-dependent.
\end{comment}

\clearpage

\onecolumngrid
\begin{center}
\textbf{\large Supplementary Materials for ``Direction reversal of non-Hermitian skin effect via coherent coupling''}\end{center}
\setcounter{equation}{0}
\setcounter{figure}{0}
\renewcommand{\theequation}{S\arabic{equation}}
\renewcommand{\thefigure}{S\arabic{figure}}
\renewcommand{\cite}[1]{\citep{#1}}

\section{Perturbative GBZ solution for two coupled Hatano-Nelson chains}
To quantitatively unveil the reversed NHSE, we solve the corresponding generalized Brillouin zone (GBZ) of two coupled Hatano-Nelson chains through a perturbative calculation.
The Hamiltonian in momentum space reads
\begin{eqnarray}
h(k)=\left(
\begin{array}{cc}
2t_a\cos (k-i\alpha_a)+\mu & t_{\perp}\\
t_{\perp} & 2t_b\cos  (k-i\alpha_b)-\mu
\end{array}
\right).
\end{eqnarray}
For large $t_{\perp}$, we may take $h_0=t_{\perp}\sigma_x$ as the unperturbed Hamiltonian, and the rest terms $h'(k)=h(k)-h_0$ as perturbations.
The first-order perturbation of the energies are given by $$\delta E_\pm(k)=\langle\psi_{\pm} |h'(k)| \psi_{\pm}\rangle$$
with $\psi_{\pm}$ the two eigenvectors of $h_0$.
Here we have $\delta E_+=\delta E_-:=\delta E$, and the two branches of eigenmodes are distinguished by their unperturbed eigenenergies $E^0_{\pm} =\pm t_{\perp}$ plus the perturbation, i.e. $$E_\pm(k)\approx  \delta E(k) \pm t_{\perp}.$$

To obtain the GBZ of our model, we introduce a complex deformation of the crystal momentum, $k\rightarrow k+i\kappa$.
%Here $\kappa$ is $k$-dependent in general, but we omit it in the notation for simplicity.
According to the non-Bloch band theory \cite{yao2018edge,yokomizo2019non}, the GBZ is given by certain values of $\kappa$ (possibly $k$-dependent) so that each pair of eigenmodes with different $k$ have the same eigenenergies.
Note that in our model, the $k$-dependency of eigenenergies appear only in the first-order perturbation $\delta E(k)$, which is the same for $E_\pm(k)$. It means that the two bands are described by the same GBZ, determined solely by $\delta E(k)$ with the complex deformation of $k$.
With some straightforward calculation, we obtain
\begin{eqnarray}
&&\delta E(k+i\kappa)=f_r\cos k+if_i\sin k,\label{eq:delta_E}\\
&&f_r=\left(t_a\frac{e^{-\kappa}e^{\alpha_a}+e^{\kappa}e^{-\alpha_a}}{2}+t_b\frac{e^{-\kappa}e^{\alpha_b}+e^{\kappa}e^{-\alpha_b}}{2}\right),\nonumber\\
&&f_i=\left(t_a\frac{e^{-\kappa}e^{\alpha_a}-e^{\kappa}e^{-\alpha_a}}{2}+t_b\frac{e^{-\kappa}e^{\alpha_b}-e^{\kappa}e^{-\alpha_b}}{2}\right).
\end{eqnarray}
%Note that here we have consider a complex deformation of the crystal momentum, $k\rightarrow k+i\kappa$ to find the GBZ of the system. Here $\kappa$ is $k$-dependent in general, but we omit it in the notation for simplicity.
%According to the non-Bloch band theory \blue{[cite]}, the GBZ is given by certain values of $\kappa$ so that each pair of eigenmodes with different $k$ have the same eigenenergies.
We can see that $\delta E(k+i\kappa)$ has its real part proportional to $\cos k$, i.e. taking the same value at $\pm k$,
and its imaginary part proportional to $\sin k$, i.e. taking the same value at $\pi/2\pm k$. Therefore, we may have $\delta E(k+i\kappa)$ taking the same value at different $k$ only when $f_rf_i=0$ \cite{li2019geometric},
which leads to
\begin{eqnarray}
e^{4\kappa}=\left(\frac{t_a e^{\alpha_a}+t_be^{\alpha_b}}{t_a e^{-\alpha_a}+t_be^{-\alpha_b}}\right)^2,~~
\kappa=\ln \sqrt{\bigg|\frac{t_a e^{\alpha_a}+t_be^{\alpha_b}}{t_a e^{-\alpha_a}+t_be^{-\alpha_b}}\bigg|}.\label{eq:kappa_two_chain}
\end{eqnarray}
For $\alpha_{a,b}>0$, the direction reversal of NHSE occurs when $\kappa<0$, which leads to the same result as that in the main text.

%Now it is clear that for certain parameters satisfying $t_at_b<0$, $\kappa$ may take negative values, meaning that the eigenmodes in the strongly coupled limit accumulate toward the opposite direction as that of the non-reciprocal accumulation of the uncoupled Hatano-Nelson chain.
%Specifically, the transition between normal and reversed skin localization occurs when $\kappa=0$, meaning that
%\begin{eqnarray}
%\frac{t_a}{-t_b}\in(\frac{e^{\alpha_b}-e^{-\alpha_b}}{e^{\alpha_a}-e^{-\alpha_a}},\frac{e^{\alpha_b}+e^{-\alpha_b}}{e^{\alpha_a}+e^{-\alpha_a}}),%\label{eq:boundary}
%\end{eqnarray}
%as shown by the solid lines in Fig. \ref{fig:fig2}(e). It is seen that these transition boundaries match well with the numerical results where the average directional IPR $\bar{I}_d$ changes its sign.
%\begin{figure}
%\includegraphics[width=1\linewidth]{phase_diagram.pdf}
%\caption{Phase diagram regarding the inverse localization length $\kappa$ in the large $t_{\perp}$ limit, obtained from Eq. %\eqref{eq:kappa_two_chain}.
%Positive and negative values $\kappa$ correspond to opposite directions of the non-reciprocal accumulation, with the phase boundaries at $\kappa=0$ (black lines) given by Eq. \eqref{eq:boundary}. Other parameters are $\alpha_a=0.5$ and $\alpha_b=0.2$.}
%\label{fig:diagram_large_v}
%\end{figure}

\section{extensions of the two coupled Hatano-Nelson chains}
\subsection{Energy offset between the two chains}
\begin{figure}[ht]
\includegraphics[width=0.8\linewidth]{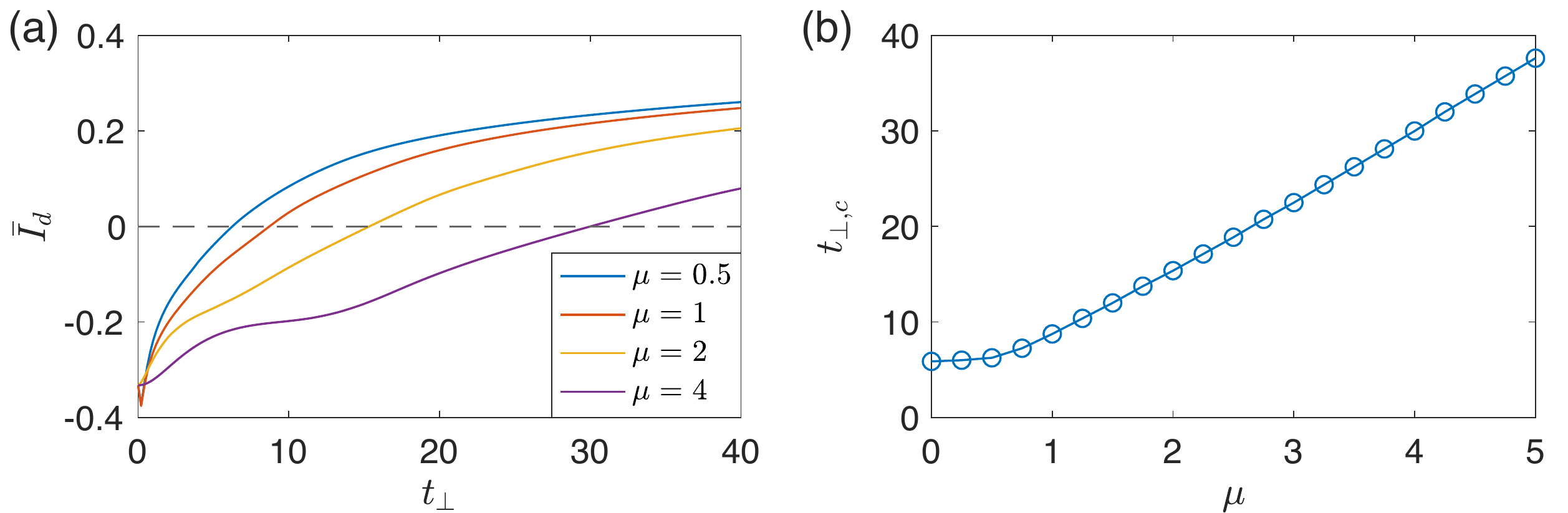}
\caption{(a) The average directional inverse participation ratio (dIPR) over all eigenmodes of two coupled Hatano-Nelson chains, for several different values of $\mu$.
See Eq. (3) in the main text for the definition of dIPR.
The transition between normal and reversed NHSE can be identified with $\bar{I}_d=0$, where the eigenmodes have a balanced distribution toward the two ends of the system.
(b) The critical value of the interchain couplings, $t_{\perp,c}$, given by $\bar{I}_d(t_{\perp,c})=0$, i.e. the crossing between the solid and dash lines in (a).
Other parameters are $t_a=0.75$, $t_b=-1$, $\alpha_a=0.5$, $\alpha_b=0.2$.
}
\label{fig:mu}
\end{figure}

The model of two coupled Hatano-Nelson chains considered in the main text contains an energy offset between the two chains, labeled as $\mu_a=-\mu_b=\mu$, which does not appear in the first-order perturbation correction.
Physically, $\mu$ tends to separate the two chains, making it more difficult for their eigensolutions to hybridize with each other.
Thus the required $t_\perp$ for entering the regime of reversed NHSE shall increase with $\mu$.
As seen in Fig. \ref{fig:mu}, the critical value of $t_\perp$ for the system to enter the regime of reversed NHSE, denoted as $t_{\perp,c}$, exhibits a roughly linear dependence on $\mu$ for $\mu\gtrsim1$.

\subsection{Off-diagonal interchain couplings}

\begin{figure}
\includegraphics[width=0.8\linewidth]{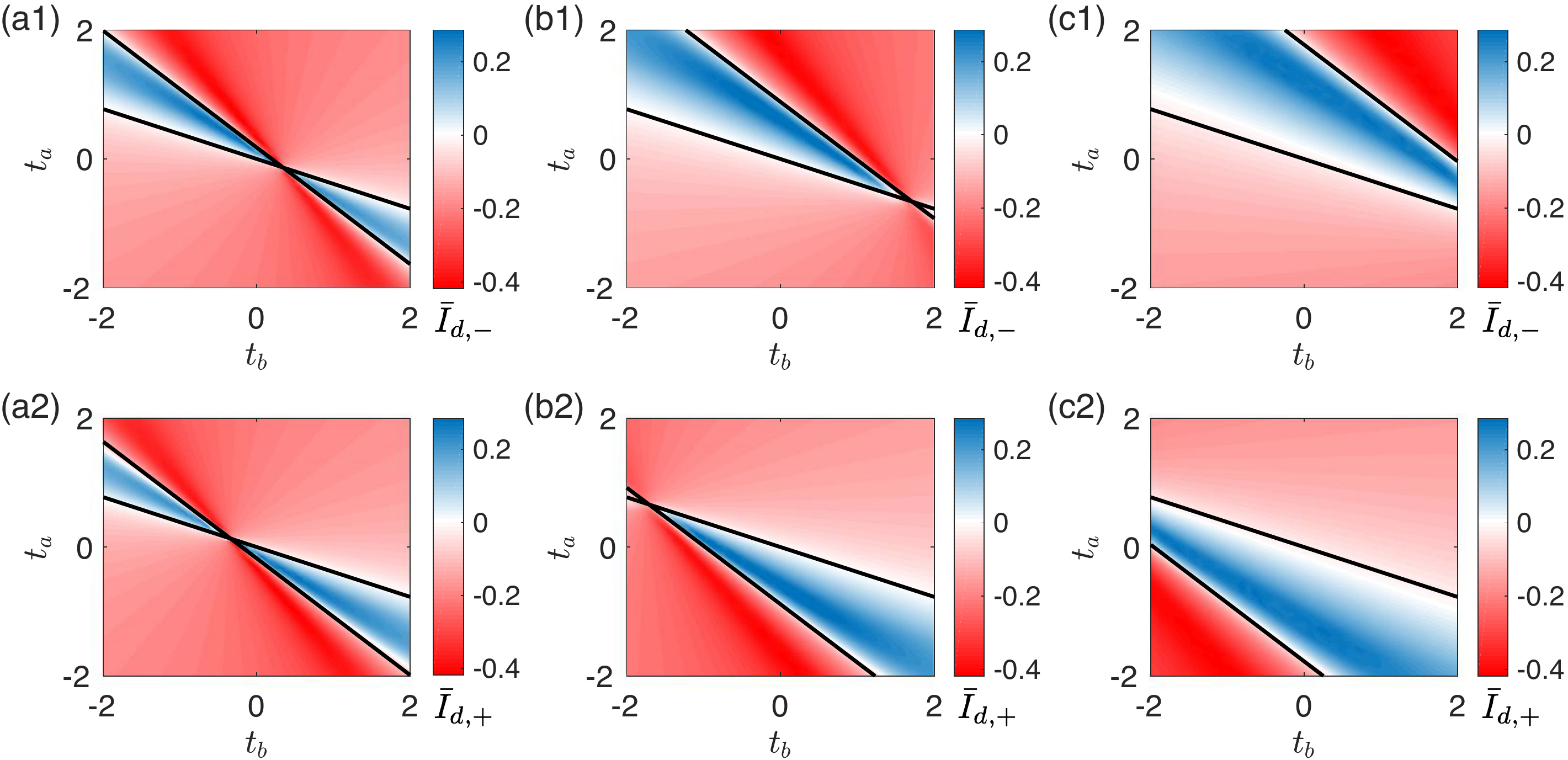}
\caption{Phase diagrams with color indicating the directional IPR $\bar{I}_{d,\pm}$ at $t_\perp=30$ with additional off-diagonal interchain couplings (a) $t_2=0.2$, (b) $t_2=1$, and (c) $t_2=2$. Black lines are obtained from the perturbation results of Eqs. \eqref{eq:boundary2} and \eqref{eq:boundary3}.
Other parameters are $\alpha_a=0.5$, $\alpha_b=0.2$, and $\mu_a=-\mu_b=0.5$.
}
\label{fig:t2}
\end{figure}
In the coupled Hatano-Nelson chains discussed in the main text, the direction reversal of NHSE occurs only when $t_a$ and $t_b$ take opposite signs, which is not a necessary condition in more general scenarios.
To see this, here we consider an example with extra off-diagonal couplings added to the model, described by
\begin{eqnarray}
\hat{H}_2=\sum_{x}\frac{t_2}{2}\hat{b}^\dagger_{x}\hat{a}_{x+1}+\frac{t_2}{2}\hat{b}^\dagger_{x}\hat{a}_{x-1}+{\rm h.c.}
\end{eqnarray}
with h.c. denoting the Hermitian conjugate. The Hamiltonian of the system in momentum space now reads
\begin{eqnarray}
h(k)=\left(
\begin{array}{cc}
2t_a\cos (k-i\alpha_a)+\mu & t_{\perp}+t_2\cos k\\
t_{\perp}+t_2\cos k & 2t_b\cos  (k-i\alpha_b)-\mu
\end{array}
\right).
\end{eqnarray}
Similar to the previous case, here we take the term of $t_\perp$ as the unperturbed Hamiltonian.
The first-order perturbation is now given by
\begin{eqnarray}
\delta E^{\pm}=t_a\cos (k-i\alpha_a)+t_b\cos (k-i\alpha_b)\pm t_2\cos k.
\end{eqnarray}
With the complex deformation $k\rightarrow k+i\kappa$, we have
\begin{eqnarray}
&&\delta E^\pm=f_r^{\pm}\cos k+if_i^{\pm}\sin k,\label{eq:delta_E_t2}\\
&&f_r^{\pm}=\left(t_a\frac{e^{-\kappa}e^{\alpha_a}+e^{\kappa}e^{-\alpha_a}}{2}+t_b\frac{e^{-\kappa}e^{\alpha_b}+e^{\kappa}e^{-\alpha_b}}{2}
\pm t_2\frac{e^{-\kappa}+e^{\kappa}}{2}\right),\nonumber\\
&&f_i^{\pm}=\left(t_a\frac{e^{-\kappa}e^{\alpha_a}-e^{\kappa}e^{-\alpha_a}}{2}+t_b\frac{e^{-\kappa}e^{\alpha_b}-e^{\kappa}e^{-\alpha_b}}{2}
\pm t_2\frac{e^{-\kappa}-e^{\kappa}}{2}\right).\nonumber\\
\end{eqnarray}
Note that here the two bands have different dependencies on $k$, therefore we shall also rewrite $\kappa$ as $\kappa_\pm$ for the two bands.
Requiring $f_r^{\pm}f_i^{\pm}=0$, we obtained
\begin{eqnarray}
e^{4\kappa_\pm}=\left(\frac{t_a e^{\alpha_a}+t_be^{\alpha_b}\pm t_2}{t_a e^{-\alpha_a}+t_be^{-\alpha_b}\pm t_2}\right)^2,\nonumber\\
\kappa_\pm=\ln \sqrt{\bigg|\frac{t_a e^{\alpha_a}+t_be^{\alpha_b}\pm t_2}{t_a e^{-\alpha_a}+t_be^{-\alpha_b}\pm t_2}\bigg|}.\label{eq:kappa_two_chain_t2}
\end{eqnarray}
The phase boundaries between normal and reversed NHSE for each band are given by $\kappa_+=0$, which leads to
\begin{eqnarray}
t_a=\frac{-t_b(e^{\alpha_b}-e^{-\alpha_b})}{e^{\alpha_a}-e^{-\alpha_a}}~{\rm and}~t_a=\frac{-t_b(e^{\alpha_b}+e^{-\alpha_b})-2t_2}{e^{\alpha_a}+e^{-\alpha_a}},\label{eq:boundary2}
\end{eqnarray}
and $\kappa_-=0$, which leads to
\begin{eqnarray}
t_a=\frac{-t_b(e^{\alpha_b}-e^{-\alpha_b})}{e^{\alpha_a}-e^{-\alpha_a}}~{\rm and}~t_a=\frac{-t_b(e^{\alpha_b}+e^{-\alpha_b})+2t_2}{e^{\alpha_a}+e^{-\alpha_a}}.\label{eq:boundary3}
\end{eqnarray}

The results for several different values of $t_2$ are shown in Fig. \ref{fig:t2}, together with the average directional IPR of the two bands, denoted as $\bar{I}_{d,\pm}$ respectively.
From this result, we can see that there are two interesting consequences due to the presence of $t_2$.

(i) Since now $\kappa_+\neq\kappa_-$, it is possible to have the reversed non-reciprocal accumulation in only one band, i.e. for exactly half of the eigenmodes. In this way, the normal and reversed non-reciprocal accumulations are separated by a large energy gap as the two bands have eigenenergies around $\pm t_\perp$ respectively.

(ii) It is now also possible to have the reversed NHSE even when $t_a$ and $t_b$ take the same sign, which is impossible for the case with only $t_\perp$ (i.e. $t_2=0$). This is because $t_2$ takes different signs in determining $\kappa_\pm$ for the two bands, and it is possible to have one of $\kappa_\pm$ changing sign by tuning $t_2$. For example, in Fig. \ref{fig:t2}(c) with $t_2=2$, the reversed NHSE (denoted by blue color) is seen when $t_{a,b}>0$ for ``$-$" band, and when $t_{a,b}<0$ for ``$+$" band.

\section{Details of the two-chain quantum walk}
The Floquet operators governing the quantum walk in the main text,
\begin{eqnarray}
U_0&=&R(\theta_1) S_2 R(\theta_2+\theta_3) M  R(\theta_2+\theta_3) S_1 R(\theta_1),\\
U&=&R(\theta_1) S_2 R(\theta_2) S_4 R(\theta_3) M  R(\theta_3) S_3R(\theta_2) S_1 R(\theta_1),
\end{eqnarray}
are explicitly given by 
\begin{eqnarray}
R(\theta)=\sum_{x=-N}^N\sum_{s=a,b}|s,x\rangle\langle s,x|\otimes e^{-i\lambda_s\theta\sigma_y/2},~~M=\sum_{x=-N}^N\sum_{s=a,b}|s,x\rangle\langle s,x|\otimes\left(|\downarrow\rangle\langle\downarrow|+e^{-\alpha_s}|\uparrow\rangle\langle\uparrow|\right),\nonumber
\end{eqnarray}
\begin{eqnarray}
S_1=\sum_{x=-N}^N\sum_{s=a,b}\left(|s,x\rangle\langle s,x|\otimes|\downarrow\rangle\langle \downarrow|+|s,x+1\rangle\langle s,x|\otimes|\uparrow\rangle\langle \uparrow|\right),~~
S_2=\sum_{x=-N}^N\sum_{s=a,b}\left(|s,x-1\rangle\langle s,x|\otimes|\downarrow\rangle\langle \downarrow|+|s,x\rangle\langle s,x|\otimes|\uparrow\rangle\langle \uparrow|\right),\nonumber
\end{eqnarray}
\begin{eqnarray}
S_3=\sum_{x=-N}^N\sum_{s,\bar{s}=a,b}\left(|s,x\rangle\langle \bar{s},x|\otimes|\uparrow\rangle\langle \uparrow|+|s,x\rangle\langle s,x|\otimes|\downarrow\rangle\langle \downarrow|\right),~~S_4=\sum_{x=-N}^N\sum_{s,\bar{s}=a,b}\left(|s,x\rangle\langle \bar{s},x|\otimes|\downarrow\rangle\langle \downarrow|
+|s,x\rangle\langle s,x|\otimes|\uparrow\rangle\langle \uparrow|\right).\nonumber
\end{eqnarray}
with $s=a,b$ denoting the two chains, $x$ the site index, and $\lambda_a=1$ and $\lambda_b=-1$.

\section{Quantum walk in a system with two unit cells}
%Experimentally it is difficult to realize a large system in quantum simulations. For the NHSE, it is a pumping effect toward the boundary of the chains, so it may becomes subtle in small systems. That is, in small systems, it will be difficult to distinguish the boundaries and the bulk. So the question is, can we still observe something to justify our discovery in small systems?
Since the NHSE describes a unidirectional transport of the (quasi-)particles or wave packets, direction reversal of NHSE in principle can manifest itself in very small systems, as long as a direction can be defined.
Specifically, we consider the quantum walk discussed in the main text in a system with the position $x$ only takes $1$ or $2$, i.e. $4$ qubits for the two chains. To distinguish the left- and right- propagations, the system must be under the ``OBC", where
%In our case we need two coupled spin-1/2 chains, so the lower limit is $4$ qubits: $2$ for each chain. And we may need an extra one as a source for the non-Hermitian gain/loss. In this case, there is no any difference between the boundaries and the bulk, yet we can still talk about pumping toward left or right, i.e. from $x=2$ to $x=1$ or the otherwise around.
%\blue{Note that in the following results, the non-Hermitian term $M$ is redefined as $$M=\sum_{x}\sum_{s=a,b}|s,x\rangle\langle s,x|\left(|\downarrow\rangle\langle\downarrow|+e^{-\alpha_s}|\uparrow\rangle\langle\uparrow|\right),$$ i.e. with loss on spin-up and no any gain, for the sack of experimental implementation. }
%Note that under the OBC,
the shift operators $S_1$ and $S_2$ are also non-unitary. That is, $S_1$ will eliminate the (quasi-)particle at $|s,2\rangle\otimes|\uparrow\rangle$, since $x=2+1$ is not included in the system. For the same reason, $S_2$ will eliminate the amplitude at $|s,1\rangle\otimes|\downarrow\rangle$. Therefore in such a small system, whether $S_1$ or $S_2$ acts first in the Floquet operator may also affect the results. To this end, we also consider another two sets of Floquet operators,
\begin{eqnarray}
\bar{U}&=&R(\theta_1) S_1 R(\theta_2) S_4 R(\theta_3) M  R(\theta_3) S_3R(\theta_2) S_2 R(\theta_1),\nonumber\\
\bar{U}_0&=&R(\theta_1) S_1 R(\theta_2+\theta_3)  M  R(\theta_2+\theta_3) S_2 R(\theta_1),\nonumber\\
\label{eq:floquet0_bar}
\end{eqnarray}
where the roles of $S_1$ and $S_2$ are exchanged when compared with $U$ and $U_0$.
%Here, we explicitly define the operators $S_{1,2,3,4}$:
%\begin{eqnarray}
%S_1&=&\sum_{x=-N}^N\sum_{s=a,b}\left(|s,x\rangle\langle s,x|\otimes|\downarrow\rangle\langle \downarrow|+|s,x+1\rangle\langle s,x|\otimes|\uparrow\rangle\langle \uparrow|\right),\nonumber\\
%S_2&=&\sum_{x=-N}^N\sum_{s=a,b}\left(|s,x-1\rangle\langle s,x|\otimes|\downarrow\rangle\langle \downarrow|+|s,x\rangle\langle s,x|\otimes|\uparrow\rangle\langle \uparrow|\right),\nonumber\\
%S_3&=&\sum_{x=-N}^N\sum_{s,\bar{s}=a,b}\left(|s,x\rangle\langle \bar{s},x|\otimes|\uparrow\rangle\langle \uparrow|+|s,x\rangle\langle s,x|\otimes|\downarrow\rangle\langle \downarrow|\right),\nonumber\\
%S_4&=&\sum_{x=-N}^N\sum_{s,\bar{s}=a,b}\left(|s,x\rangle\langle \bar{s},x|\otimes|\downarrow\rangle\langle \downarrow|+|s,x\rangle\langle s,x|\otimes|\uparrow\rangle\langle \uparrow|\right),\nonumber
%\end{eqnarray}
%with $s\neq \bar{s}$.

\begin{figure}
\includegraphics[width=0.5\linewidth]{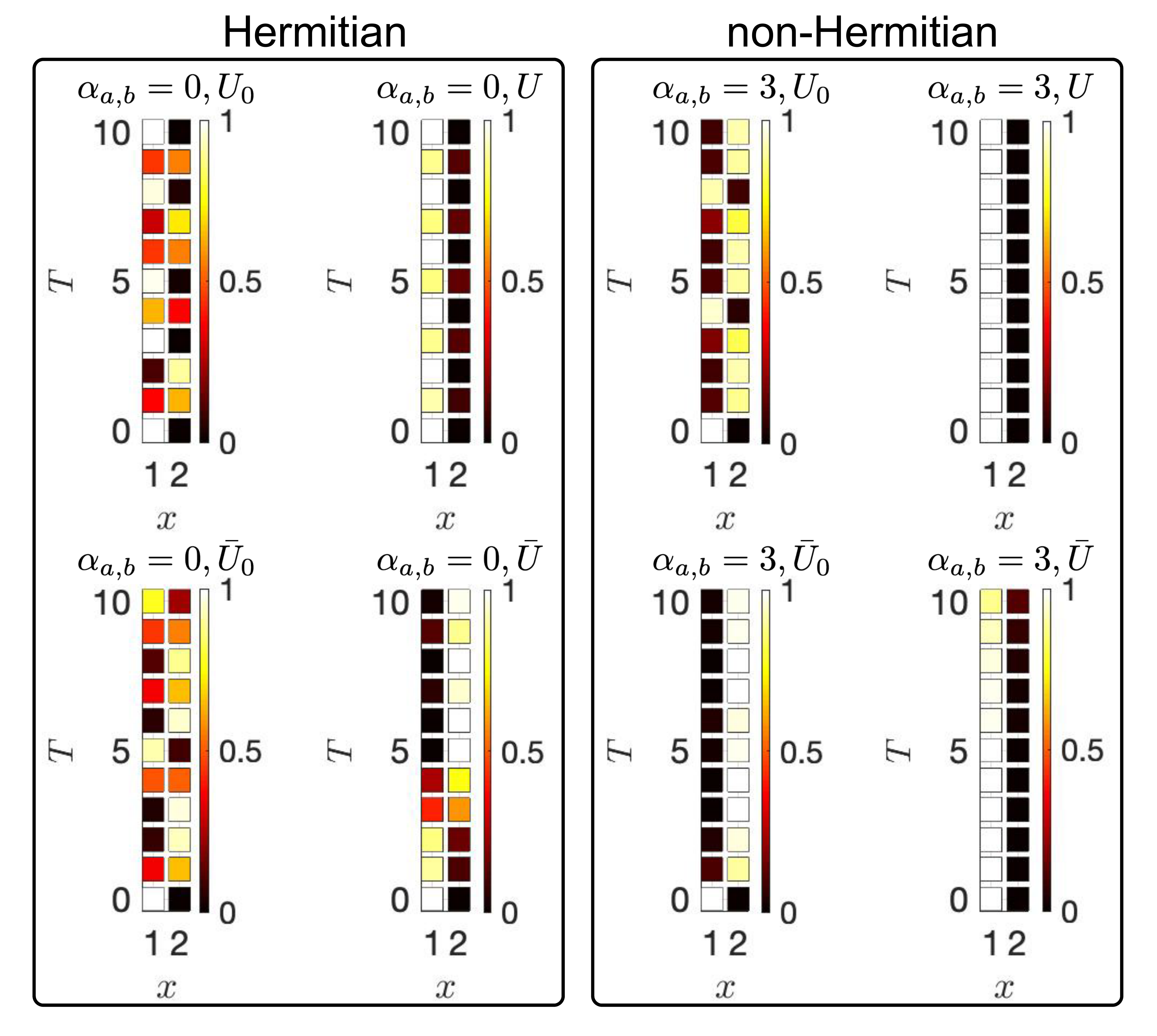}
\caption{
The normalized spatial distribution $\rho(x)$ [or $\rho_0(x)$, $\bar{\rho}(x)$, $\bar{\rho}_0(x)$] for a quantum walk with two unit cells, i.e. $x=1$ or $2$.
The initial state is prepared as $\Psi_{\rm ini}=|a,1\rangle\otimes|\uparrow\rangle$.
From left to right, the results are obtained with the system being (i) Hermitian and decoupled; (ii) Hermitian and coupled; (iii) non-Hermitian and decoupled, (iv) non-Hermitian and coupled.
The parameters are $\alpha_a=\alpha_b=0$ and $\alpha_a=\alpha_b=3$ for Hermitian and non-Hermitian cases respectively.
}
\label{fig:QW_size2_a1}
\end{figure}

By comparing the dynamical evolutions governed by these four Floquet operators ($U$, $U_0$, $\bar{U}$, and $\bar{U}_0$) with or without the non-Hermitian loss,
%i.e. for nonzero and zero $\alpha_{a,b}$,
the direction reversal of NHSE induced by the interchain couplings can be justified even in this two-unit-cell system, as demonstrated in Fig. \ref{fig:QW_size2_a1}.
%In Fig. \ref{fig:QW_size2_a1} we illustrate the results for quantum walks governed by these four Floquet operators, each with or without the non-Hermitian loss, i.e. for nonzero and zero $\alpha_{a,b}$.
Below we summarize the results. Starting from the left of Fig. \ref{fig:QW_size2_a1},

(i) in the first column, we choose $\alpha_{a,b}=0$, and consider the quantum walks without the interchain couplings, governed by $U_0$ and $\bar{U}_0$ respectively. There is no non-reciprocal pumping since there is no gain or loss in the system.

(ii) in the second column, we choose $\alpha_{a,b}=0$, and consider the quantum walks with the interchain couplings, governed by $U$ and $\bar{U}$ respectively. Even without any gain or loss in the system, we still observe some non-reciprocal accumulations. However, we can see that $U$ and $\bar{U}$ (with $S_{1,2}$ exchanging their roles)  induce different accumulating directions, showing that this is not the NHSE, but a consequence of the boundary effect induced by $S_1$ and $S_2$.

(iii) in the third column, we choose $\alpha_{a,b}=3$, and consider the same quantum walks as that of the first column.
We see a clear non-reciprocal accumulation from $x=1$ to $x=2$, for both Floquet operators. Therefore we conclude that this non-reciprocal accumulation reflects the NHSE, rather than the boundary effect induced by $S_1$ and $S_2$.

(iv) in the last column, we choose $\alpha_{a,b}=3$, and consider the same quantum walks as that of the second column.
A clear non-reciprocal accumulation is seen from $x=2$ to $x=1$, also for both Floquet operators. Compared with (iii), we can see that the direction of population accumulation is reversed when introducing $S_3$ and $S_4$, representing the direction reversal of NHSE induced by interchain couplings.

%Finally, we note that here the initial state is prepared at $\Psi_{\rm ini}=|a,1\rangle\otimes|\uparrow\rangle$, and qualitatively the same results can be obtained also for an initial state prepared at $x=2$.

\begin{figure}
\includegraphics[width=0.5\linewidth]{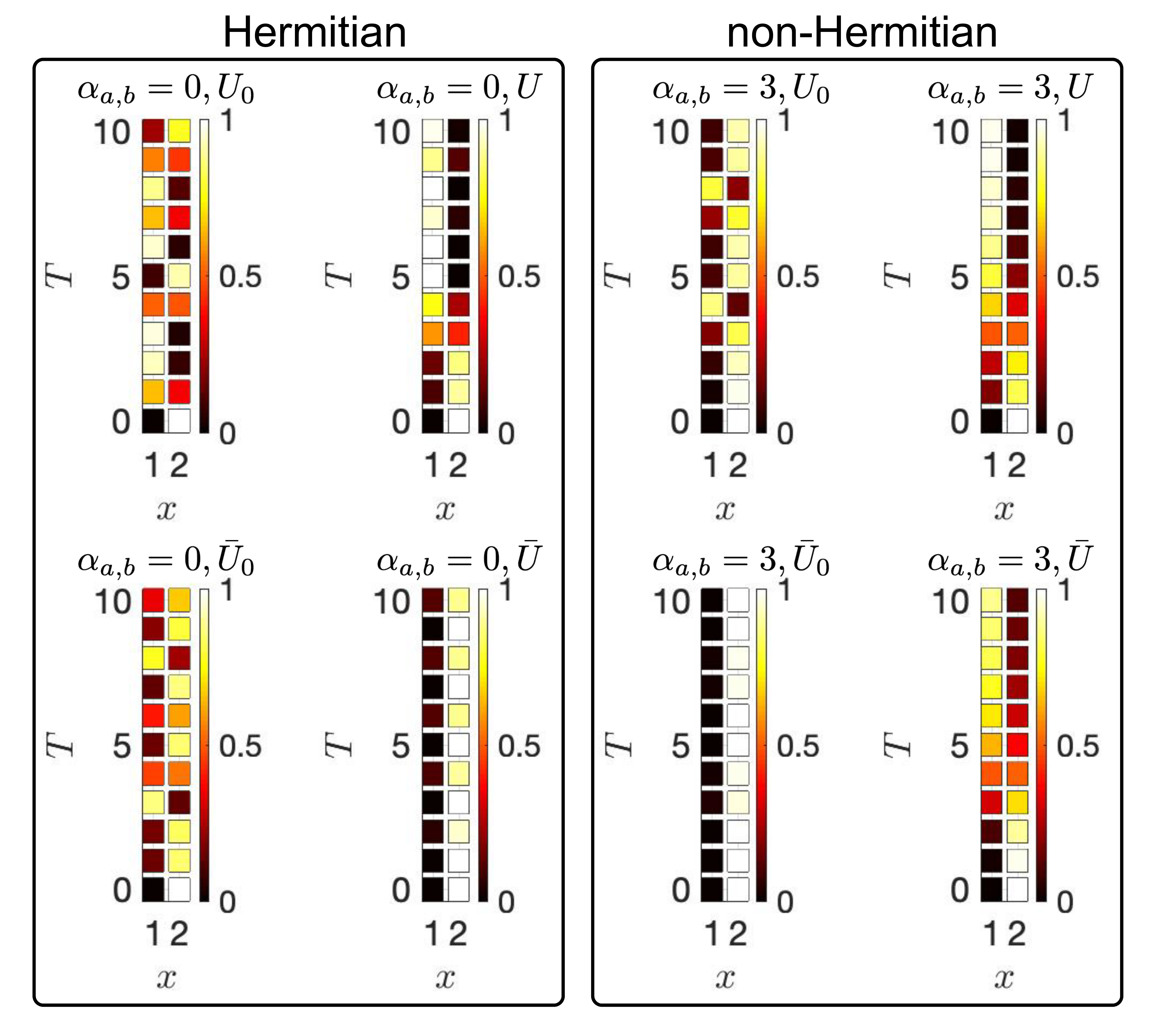}
\caption{
The normalized spatial distribution $\rho(x)$ [or $\rho_0(x)$, $\bar{\rho}(x)$, $\bar{\rho}_0(x)$] for a quantum walk with two unit cells, i.e. $x=1$ or $2$.
The initial state is prepared as $\Psi_{\rm ini}=|a,2\rangle\otimes|\uparrow\rangle$, which is different from that in Fig. \ref{fig:QW_size2_a1}.
From left to right, the results are obtained with (i) Hermitian and decoupled; (ii) Hermitian and coupled; (iii) non-Hermitian and decoupled, (iv) non-Hermitian and coupled. Hermitian and non-Hermitian cases correspond to $\alpha_a=\alpha_b=0$ and $\alpha_a=\alpha_b=3$ respectively.
}
\label{fig:QW_size2_a2}
\end{figure}

In Fig.~\ref{fig:QW_size2_a2}, keeping parameters the same with those in Fig. \ref{fig:QW_size2_a1}, we present the results with a different initial state $\Psi_{\rm ini}=|a,2\rangle\otimes|\uparrow\rangle$ and we observe the same phenomena as shown in Fig. \ref{fig:QW_size2_a1}.

\bibliography{references.bib}
\end{document}